\documentclass[12pt]{article}

\usepackage{amsmath,amssymb,graphicx,color,array,cite}
\numberwithin{equation}{section}

\evensidemargin \oddsidemargin

\begin{document}

\begin{titlepage}

\renewcommand{\thefootnote}{\fnsymbol{footnote}}

\hfill\parbox{35mm}{hep-th/0409045 \\ KEK-TH-975}

\vspace{15mm}
\baselineskip 9mm
\begin{center}
  {\Large \bf Membrane Fuzzy Sphere Dynamics \\
    in Plane-Wave Matrix Model}
\end{center}

\baselineskip 6mm
\vspace{10mm}
\begin{center}
  Hyeonjoon Shin$^a$\footnote{\tt hshin@newton.skku.ac.kr} 
  and Kentaroh Yoshida$^b$\footnote{\tt
    kyoshida@post.kek.jp} \\[10mm] 
  {\sl $^a$BK 21 Physics Research
    Division and Institute of Basic Science\\ 
    Sungkyunkwan University,
    Suwon 440-746, South Korea \\[3mm] 
    $^b$Theory Division, High Energy
    Accelerator Research Organization (KEK)\\ 
    Tsukuba, Ibaraki 305-0801, Japan}
\end{center}

\thispagestyle{empty}

\vfill
\begin{center}
{\bf Abstract}
\end{center}
\noindent
In plane-wave matrix model, the membrane fuzzy sphere extended in the
$SO(3)$ symmetric space is allowed to have periodic motion on a
sub-plane in the $SO(6)$ symmetric space.  We consider a background
configuration composed of two such fuzzy spheres moving on the same
sub-plane and the one-loop quantum corrections to it.  The one-loop
effective action describing the fuzzy sphere interaction is computed
up to the sub-leading order in the limit that the mean distance $r$
between two fuzzy spheres is very large.  We show that the leading
order interaction is of the $1/r^7$ type and thus the membrane fuzzy
spheres interpreted as giant gravitons really behave as gravitons.
\\ [5mm] Keywords : pp-wave, Matrix model, Fuzzy sphere 
\\ PACS numbers : 11.25.-w, 11.27.+d, 12.60.Jv

\vspace{5mm}
\end{titlepage}

\baselineskip 6.6mm
\renewcommand{\thefootnote}{\arabic{footnote}}
\setcounter{footnote}{0}

\section{Introduction and Summary}

The plane-wave matrix model \cite{Berenstein:2002jq} is a microscopic
description of the discrete light cone quantized (DLCQ) M-theory in
the eleven-dimensional $pp$-wave or plane-wave background
\cite{Kowalski-Glikman:1984wv}, which is $SO(3) \times SO(6)$
symmetric and given by
\begin{align}
ds^2 &= -2 dx^+ dx^- 
    - \left( \sum^3_{i=1} \left( \frac{\mu}{3} \right)^2 (x^i)^2
            +\sum^9_{a=4} \left( \frac{\mu}{6} \right)^2 (x^a)^2
      \right) (dx^+)^2
    + \sum^9_{I=1} (dx^I)^2 ~,
     \notag \\
F_{+123} &= \mu ~,
\label{pp}
\end{align}
with the index notation $I=(i,a)$. It is now well known that this
background is maximally supersymmetric and the limiting case of the
eleven-dimensional $AdS$ type geometries \cite{Blau:2002dy}.  Due to
the effect of the $++$ component of the metric and the presence of the
four-form field strength, the plane-wave matrix model has some $\mu$
dependent terms, which make the difference between the usual flat
space matrix model \cite{Banks:1997vh} and the plane-wave one.

Compared to the flat matrix model, the plane-wave matrix model has
some peculiar properties.  One of them is that there are various
supersymmetric vacuum structures classified by the $SU(2)$ algebra,
which may be identified as collections of membrane fuzzy spheres
\cite{Dasgupta:2002hx}.  Motivated by this classification, there have
been further studies on the vacuum structure especially related to the
protected multiplet \cite{Kim:2002if}, and various supersymmetric
extended objects allowed by the plane-wave matrix model have been
searched and studied \cite{Bak:2002rq,Sugiyama:2002rs,Hyun:2002cm,
  Sugiyama:2002bw,Mikhailov:2002wx,Park:2002cb,Maldacena:2002rb,
  Yee:2003ge,Kim:2002cr,Hyun:2002fk,Freedman:2003kb}.  There also have
been studies on the thermodynamic properties of vacua
\cite{Huang:2003gw}.

The basic ingredient for vacua of plane-wave matrix model is the
membrane fuzzy sphere which is interpreted as giant graviton.  It
preserves the full 16 supersymmetries of the plane-wave matrix model
and exists even when the matrices have finite size $N$.  Recently,
based on the known properties of the membrane fuzzy sphere, we have
tried to investigate the dynamical aspects of the plane-wave matrix
model at the one-loop level and shown that two fuzzy spheres taking
circular motion on a sub-plane of $SO(6)$ symmetric transverse space do
not interact \cite{Shin:2003np}.  Single fuzzy sphere sitting at the
origin of $SO(3)$ symmetric space and rotating in $SO(6)$ symmetric
one is known to be a BPS object preserving 8 supersymmetries
\cite{Park:2002cb}.  Our previous result implies that the
configuration composed of two or more such fuzzy spheres rotating on
the same sub-plane is also supersymmetric.  This intriguing result
however does not tell us about the dynamics of fuzzy spheres or giant
gravitons.

In order to investigate the fuzzy sphere dynamics in plane-wave matrix
model, it is then necessary to begin with a background configuration
of fuzzy spheres which breaks supersymmetry explicitly.  In this
paper, we consider such a background and its one-loop quantum
corrections.  The resulting effective action describes the interaction
between two fuzzy spheres, which respects the basic dynamical aspects
of the plane-wave matrix model.

\begin{figure}
\begin{center}
 \includegraphics[scale=.6]{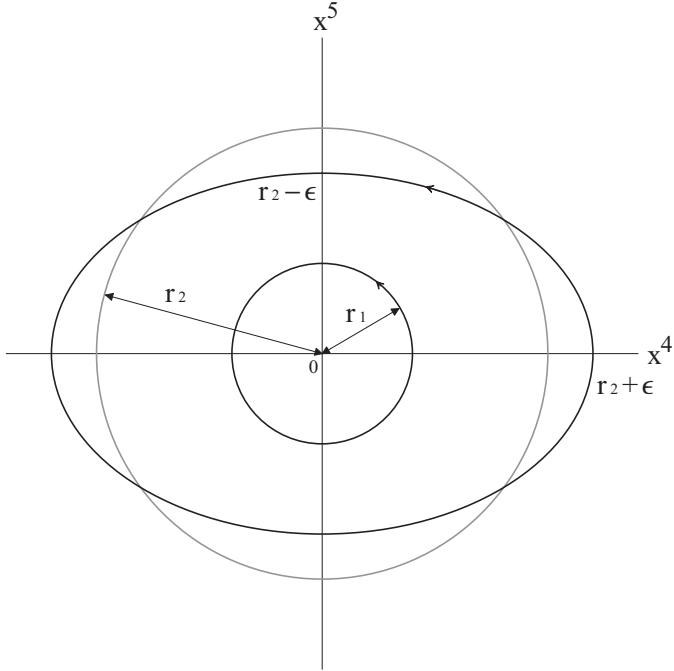}
\end{center}
\caption{Configuration of two membrane fuzzy spheres moving in the 
  $x^4$-$x^5$ plane.  Since they span in the $SO(3)$ symmetric space,
  they are actually points in this plane. The first fuzzy sphere is in
  circular motion with the radius $r_1$.  The second one takes an
  elliptic motion whose major and minor semi-axis are $r_2+\epsilon$
  and $r_2-\epsilon$ respectively.}
\label{config}
\end{figure}

In the plane-wave matrix model, the static and supersymmetric fuzzy
sphere is specified by the fine $N$-dimensional irreducible
representation of $SU(2)$ and spans in the $SO(3)$ symmetric space.
From now on, let us call $N$ the size of the fuzzy sphere.  As will be
discussed in Sec.~\ref{pp-matrix}, the equations of motion for the
matrix variables allow it to have periodic motion.  Let us consider a
fuzzy sphere which has periodic motion on a sub-plane of the $SO(6)$
symmetric space (The $x^4$-$x^5$ plane is chosen in this paper).  It
has been known that it is supersymmetric when its periodic motion is
circular, and otherwise it is not \cite{Park:2002cb}.  If we now
consider a background configuration composed of two fuzzy spheres
whose motions are circular and elliptic respectively in the
$x^4$-$x^5$ plane, then it clearly breaks all the supersymmetry.  The
fuzzy sphere configuration taken in this paper is shown in
Fig.~\ref{config} and will be described in more detail in
Sec.~\ref{bg-exp}.  We note that the sizes of two fuzzy spheres are
$N_1$ and $N_2$ respectively, and $N=N_1+N_2$.

The plane-wave matrix model expanded around the fuzzy sphere is
allowed to be studied perturbatively in the large $\mu$ limit
\cite{Dasgupta:2002hx}.  In particular, in the limit of $\mu
\rightarrow \infty$, only the quadratic action for the fluctuations
around the fuzzy sphere background is important and hence the one-loop
study gives exact results.  In Sec.~\ref{bg-exp}, this is justified
again for our background, Fig.~\ref{config}.  The path integration of
the quadratic action is performed in Sec.~\ref{path}, where only the
formal results are given.  The actual calculation for obtaining the
effective action is not easy due to the time dependent background
given by trigonometric functions.  At this stage, we need a proper
parameter in terms of which the formal expressions of path
integrations can be expanded perturbatively.  It turns out that
$\epsilon$ can be such a parameter if we assume that $\epsilon \ll
|r_2 -r_1|$, that is, the elliptic motion of the second fuzzy sphere
is taken to be almost circular.  

In Sec.~\ref{eff-int}, we present the perturbative calculation of the
effective action in powers of $\epsilon$, and find that there are
non-vanishing contributions from the $\epsilon^4$ order.  The
effective action describing the interaction between two fuzzy spheres
is given by a function of $r$, $N_1$, and $N_2$, where $r \equiv
r_2-r_1$ is the mean distance between two fuzzy spheres for one period
of motion, and is evaluated in the long distance limit ($r \gg N$).
Its explicit expression is obtained as
\begin{align}
\Gamma_{\rm eff} = \epsilon^4 \int dt \left[
 \frac{35}{2^7 \cdot 3} \frac{N_1 N_2}{r^7}
-\frac{385}{2^{11} \cdot 3^3} 
  \big[ 2 (N_1^2 + N_2^2) -1 \big] \frac{N_1 N_2}{r^9}
+  \mathcal{O} \left( \frac{1}{r^{11}} \right) \right] 
+ \mathcal{O} ( \epsilon^6 ) ~.
\end{align}
In obtaining this effective action, there are contributions from the
various sectors.  The contributions themselves are presented in an
appendix \ref{appen}.  We note that, as will be discussed in
Sec.~\ref{eff-int}, the effective action actually does not change even
when $r_1 > r_2$ and thus $r$ should be regarded as $|r|$.  From the
above effective action, we see that the leading order interaction is
given by the form of $1/r^7$.  This clearly shows that the two fuzzy
spheres interpreted as giant gravitons really behave as gravitons.

\section{Plane-wave matrix model and classical solutions}
\label{pp-matrix}

The plane-wave matrix model is basically composed of two parts.  One
part is the usual matrix model based on  eleven-dimensional flat
space-time, that is, the flat space matrix model, and another is a set
of terms reflecting the structure of the maximally supersymmetric
eleven dimensional plane-wave background, Eq. (\ref{pp}).  Its action
is
\begin{equation}
S_{pp} = S_\mathrm{flat} + S_\mu ~,
\label{pp-bmn}
\end{equation}
where each part of the action on the right hand side is given by
\begin{align}
S_\mathrm{flat} & = \int dt \mathrm{Tr} 
\left( \frac{1}{2R} D_t X^I D_t X^I + \frac{R}{4} ( [ X^I, X^J] )^2
      + i \Theta^\dagger D_t \Theta 
      - R \Theta^\dagger \gamma^I [ \Theta, X^I ]
\right) ~,
  \notag \\
S_\mu &= \int dt \mathrm{Tr}
\left( 
      -\frac{1}{2R} \left( \frac{\mu}{3} \right)^2 (X^i)^2
      -\frac{1}{2R} \left( \frac{\mu}{6} \right)^2 (X^a)^2
      - i \frac{\mu}{3} \epsilon^{ijk} X^i X^j X^k
      - i \frac{\mu}{4} \Theta^\dagger \gamma^{123} \Theta
\right) ~.
\label{o-action}
\end{align}
Here, $R$ is the radius of circle compactification along $x^-$ and
$D_t$ is the covariant derivative with the gauge field $A$,
\begin{equation}
D_t = \partial_t - i [A, \: ] ~.
\end{equation}
The matrices $\gamma^I$ are the $SO(9)$ gamma matrices and satisfy the
Clifford algebra
\begin{align}
\{ \gamma^I, \gamma^J \} = 2 \delta^{IJ} ~.
\label{so9-cliff}
\end{align}

For dealing with the problem in this paper, it is convenient to
rescale the gauge field and parameters as
\begin{equation}
A \rightarrow R A ~,~~~ 
t \rightarrow \frac{1}{R} t ~,~~~
\mu \rightarrow R \mu ~.
\end{equation}
With this rescaling, the radius parameter $R$ disappears and the
actions in Eq.~(\ref{o-action}) become
\begin{align}
S_\mathrm{flat} & = \int dt \mathrm{Tr} 
\left( \frac{1}{2} D_t X^I D_t X^I + \frac{1}{4} ( [ X^I, X^J] )^2
      + i \Theta^\dagger D_t \Theta 
      -  \Theta^\dagger \gamma^I [ \Theta, X^I ]
\right) ~,
  \notag \\
S_\mu &= \int dt \mathrm{Tr}
\left( 
      -\frac{1}{2} \left( \frac{\mu}{3} \right)^2 (X^i)^2
      -\frac{1}{2} \left( \frac{\mu}{6} \right)^2 (X^a)^2
      - i \frac{\mu}{3} \epsilon^{ijk} X^i X^j X^k
      - i \frac{\mu}{4} \Theta^\dagger \gamma^{123} \Theta
\right) ~.
\label{pp-action}
\end{align}

The possible backgrounds allowed by the plane-wave matrix model are
the classical solutions of the equations of motion for the matrix
fields.  Since the background that we are concerned about is purely
bosonic, we concentrate on solutions of the bosonic fields $X^I$.  We
would like to note that we will not consider all possible solutions
but only those relevant to our interest for the fuzzy sphere
interaction.  Then, from the rescaled action, (\ref{pp-action}), the
bosonic equations of motion are derived as
\begin{align}
\ddot{X}^i &=
   - [[ X^i, X^I],X^I] - \left( \frac{\mu}{3} \right)^2 X^i
   - i \mu \epsilon^{ijk} X^j X^k ~, 
                 \nonumber \\
\ddot{X}^a &=
    - [[ X^a, X^I],X^I] - \left( \frac{\mu}{6} \right)^2 X^a ~,
\label{eom}
\end{align}
where the over dot implies the time derivative $\partial_t$.  

Except for the trivial $X^I=0$ solution, the simplest one is the
harmonic oscillator solution;
\begin{equation}
X^i_\mathrm{osc} 
= A^i \cos \left( \frac{\mu}{3} t + \phi_i \right) 
 {\bf 1}_{N \times N} ~,~~~
X^a_\mathrm{osc} 
= A^a \cos \left( \frac{\mu}{6} t + \phi_a \right) 
 {\bf 1}_{N \times N} ~,
\label{osc}
\end{equation}
where $A^I$ and $\phi_I $ $(I=(i,a))$ are the amplitudes and phases of
oscillations respectively, and ${\bf 1}_{N \times N}$ is the $N \times
N$ unit matrix.  This oscillatory solution is special to the
plane-wave matrix model due to the presence of mass terms for $X^I$.
It should be noted that, because of the mass terms, the configuration
corresponding to the time dependent straight line motion, say $v^I t +
c^I$ with non-zero constants $v^I$ and $c^I$, is not possible as a
solution of (\ref{eom}), that is, a classical background of plane-wave
matrix model, contrary to the case of the flat space matrix model.  As
the generalization of the oscillatory solution, Eq. (\ref{osc}), we
get the solution of the form of diagonal matrix with each diagonal
element having independent amplitude and phase.

As for the non-trivial constant matrix solution, Eq. (\ref{eom})
allows the following membrane fuzzy sphere or giant graviton solution:
\begin{equation}
X^i_\mathrm{sphere} = \frac{\mu}{3} J^i ~,
\label{fuzzy}
\end{equation}
where $J^i$ satisfies the $SU(2)$ algebra,
\begin{equation}
[ J^i, J^j ] = i \epsilon^{ijk} J^k ~.
\label{su2}
\end{equation}
The reason why this solution is possible is basically that the matrix
field $X^i$ feels an extra force due to the Myers interaction which
may stabilize the oscillatory force.  The fuzzy sphere solution
$X^i_\mathrm{sphere}$ preserves the full 16 dynamical supersymmetries
of the plane-wave and hence is 1/2-BPS object.  We note that actually
there is another fuzzy sphere solution of the form $\frac{\mu}{6}
J^i$.  However, it has been shown that such solution receives quantum
corrections and hence is non-BPS object \cite{Sugiyama:2002bw}.

\section{Expansion around fuzzy sphere configuration}
\label{bg-exp}

In this section, the plane-wave matrix model is expanded around the
fuzzy sphere configuration in which we are interested.

We first consider the expansion around the general background.  For
this, let us split the matrix quantities into as follows:
\begin{equation}
X^I = B^I + Y^I ~,~~~ \Theta = F + \Psi ~,
\label{cl+qu}
\end{equation}
where $B^I$ and $F$ are the general classical background fields
satisfying the classical equations of motion, Eq. (\ref{eom}), while
$Y^I$ and $\Psi$ are the quantum fluctuations around them.  The
fermionic background $F$ is taken to vanish from now on, since we will
only consider the purely bosonic background.  The quantum fluctuations
are the fields subject to the path integration.  In taking into
account the quantum fluctuations, we should recall that the matrix
model itself is a gauge theory.  This implies that the gauge fixing
condition should be specified before proceed further.  In this paper,
we take the background field gauge which is usually chosen in the
matrix model calculation as
\begin{equation}
D_\mu^{\rm bg} A^\mu_{\rm qu} \equiv
D_t A + i [ B^I, X^I ] = 0 ~.
\label{bg-gauge}
\end{equation}
Then the corresponding gauge-fixing $S_\mathrm{GF}$ and Faddeev-Popov
ghost $S_\mathrm{FP}$ terms are given by
\begin{equation}
S_\mathrm{GF} + S_\mathrm{FP} =  \int\!dt \,{\rm Tr}
  \left(
      -  \frac{1}{2} (D_\mu^{\rm bg} A^\mu_{\rm qu} )^2 
      -  \bar{C} \partial_{t} D_t C + [B^I, \bar{C}] [X^I,\,C]
\right) ~.
\label{gf-fp}
\end{equation}

Now by inserting the decomposition of the matrix fields (\ref{cl+qu})
into Eqs.~(\ref{pp-action}) and (\ref{gf-fp}), we get the gauge fixed
plane-wave action $S$ $(\equiv S_{pp} + S_\mathrm{GF} +
S_\mathrm{FP})$ expanded around the background.  The resulting action
is read as
\begin{equation}
S =  S_0 + S_2 + S_3 + S_4 ~,
\end{equation}
where $S_n$ represents the action of order $n$ with respect to the
quantum fluctuations and, for each $n$, its expression is
\begin{align}
S_0 = \int dt \, \mathrm{Tr} \bigg[ \,
&      \frac{1}{2}(\dot{B}^I)^2  
        - \frac{1}{2} \left(\frac{\mu}{3}\right)^2 (B^i)^2 
        - \frac{1}{2} \left(\frac{\mu}{6}\right)^2 (B^a)^2 
        + \frac{1}{4}([B^I,\,B^J])^2
        - i \frac{\mu}{3} \epsilon^{ijk} B^i B^j B^k 
    \bigg] ~,
\notag \\
S_2 = \int dt \, \mathrm{Tr} \bigg[ \,
&       \frac{1}{2} ( \dot{Y}^I)^2 - 2i \dot{B}^I [A, \, Y^I] 
        + \frac{1}{2}([B^I , \, Y^J])^2 
        + [B^I , \, B^J] [Y^I , \, Y^J]
        - i \mu \epsilon^{ijk} B^i Y^j Y^k
\notag \\
&       - \frac{1}{2} \left( \frac{\mu}{3} \right)^2 (Y^i)^2 
        - \frac{1}{2} \left( \frac{\mu}{6} \right)^2 (Y^a)^2 
        + i \Psi^\dagger \dot{\Psi} 
        -  \Psi^\dagger \gamma^I [ \Psi , \, B^I ] 
        -i \frac{\mu}{4} \Psi^\dagger \gamma^{123} \Psi  
\notag \\ 
&       - \frac{1}{2} \dot{A}^2  - \frac{1}{2} ( [B^I , \, A])^2 
        + \dot{\bar{C}} \dot{C} 
        + [B^I , \, \bar{C} ] [ B^I ,\, C] \,
     \bigg] ~,
\notag \\
S_3 = \int dt \, \mathrm{Tr} \bigg[
&       - i\dot{Y}^I [ A , \, Y^I ] - [A , \, B^I] [ A, \, Y^I] 
        + [ B^I , \, Y^J] [Y^I , \, Y^J] 
        +  \Psi^\dagger [A , \, \Psi] 
\notag \\
&       -  \Psi^\dagger \gamma^I [ \Psi , \, Y^I ] 
        - i \frac{\mu}{3} \epsilon^{ijk} Y^i Y^j Y^k
        - i \dot{\bar{C}} [A , \, C] 
        +  [B^I,\, \bar{C} ] [Y^I,\,C]  \,
     \bigg] ~,
\notag \\
S_4 = \int dt \, \mathrm{Tr} \bigg[
&       - \frac{1}{2} ([A,\,Y^I])^2 + \frac{1}{4} ([Y^I,\,Y^J])^2 
     \bigg] ~.
\label{bgaction} 
\end{align}

We now set up the background configuration for the membrane fuzzy
spheres.  Since we will study the interaction of two fuzzy spheres,
the matrices representing the background have the $2 \times 2$ block
diagonal form as
\begin{equation}
B^I = \begin{pmatrix} B_{(1)}^I & 0 \\ 0 & B_{(2)}^I
      \end{pmatrix} ~,
\label{gb}
\end{equation}
where $B_{(s)}^I$ with $s=1,2$ are $N_s \times N_s$ matrices.  If we
take $B^I$ as $N \times N$ matrices, then $N = N_1 + N_2$.  

The two fuzzy spheres are taken to be static in the space where they
span, and hence represented by the classical solution, (\ref{fuzzy});
\begin{equation}
B_{(s)}^i = \frac{\mu}{3} J_{(s)}^i ~,
\label{b3}
\end{equation}
where, for each $s$, $J_{(s)}^i$ is in the $N_s$-dimensional
irreducible representation of $SU(2)$ and satisfies the $SU(2)$
algebra, Eq. (\ref{su2}).  In the $SO(6)$ symmetric transverse space,
the fuzzy spheres are regarded as point objects, of course, in a sense
of ignoring the matrix nature.  We first take a certain
two-dimensional sub-plane in the $SO(6)$ symmetric space.  The two
fuzzy spheres are then led to have periodic motion in this plane.
More precisely, the first fuzzy sphere given by the background of
$s=1$ is taken to be in circular motion with the radius $r_1$. As for
the second fuzzy sphere, we take an elliptic motion whose major and
minor semi-axis are $r_2+\epsilon$ and $r_2-\epsilon$ respectively.
Obviously, this configuration is one of the classical solutions of the
equations of motion as one can see from Eq.~(\ref{osc}).  Recalling
that the transverse space is $SO(6)$ symmetric, all the possible
choices of two-dimensional sub-plane where the periodic motion takes
place are equivalent.  Thus, without loss of generality, we can take
an arbitrary plane for the periodic motion.  In this paper, the
$x^4$-$x^5$ plane is chosen.  Then the configuration in the transverse
space is given by
\begin{gather}
B^4_{(1)} = r_1 \cos \left( \frac{\mu}{6} t \right) 
             {\bf 1}_{N_1 \times N_1}~,~~~
B^5_{(1)} = r_1 \sin \left( \frac{\mu}{6} t \right) 
             {\bf 1}_{N_1 \times N_1}~,
\notag \\
B^4_{(2)} = (r_2 + \epsilon) \cos \left( \frac{\mu}{6} t \right) 
             {\bf 1}_{N_2 \times N_2}~,~~~
B^5_{(2)} = (r_2 - \epsilon) \sin \left( \frac{\mu}{6} t \right) 
             {\bf 1}_{N_2 \times N_2}~,
\label{b6}
\end{gather}
where $r_1$, $r_2$, and $\epsilon$ are non-negative.  Eqs.~(\ref{b3})
and (\ref{b6}) compose the background configuration about which we are
concerned, and all other elements of matrices $B^I$ are set to zero. A
schematic view of the background configuration is presented in
Fig.~\ref{config}.  We would like to note that the fuzzy sphere
represented by $B^I_{(1)}$ is supersymmetric \cite{Park:2002cb}.
However, since the fuzzy sphere in elliptic motion is not
supersymmetric unless $\epsilon = 0$, the whole configuration does not
have supersymmetry.  Although it is not necessary at the present
stage, we assume that $\epsilon \ll r_2$ or $\epsilon \ll |r_2-r_1|$.
As we shall see in later sections, the actual evaluation of the
one-loop effective action requires a perturbative expansion parameter
and the parameter $\epsilon$ can play a role of such parameter.  Thus,
the second fuzzy sphere is assumed to take an elliptic motion which is
almost circular.

Actually, two fuzzy spheres may have their motions in different
sub-planes in the $SO(6)$ symmetric space.  The configuration taken
here may be the simplest one.  However, since our purpose is to
investigate the very basic dynamical aspects of fuzzy spheres which
are to be seen in every situations, it will be sufficient to consider
only the simplest one among the possible configurations which seem to
give non-trivial result.  Indeed, as we shall see, the configuration
given by Eqs.~(\ref{b3}) and (\ref{b6}) leads to the basic but
non-trivial interaction between two fuzzy spheres.

The classical value of the action for the background
configuration,(\ref{b3}) and (\ref{b6}), is simply zero;
\begin{equation}
S_0 = 0 ~.
\label{clav}
\end{equation}
In the following sections, we are going to compute the one-loop
correction to this action, that is, to the background, (\ref{b3}) and
(\ref{b6}), due to the quantum fluctuations via the path integration
of the quadratic action $S_2$, and obtain the one-loop effective
action $\Gamma_\mathrm{eff}$ or the effective potential
$V_\mathrm{eff}$.

For the justification of one-loop computation or the semi-classical
analysis, it should be made clear that $S_3$ and $S_4$ of
Eq.~(\ref{bgaction}) can be regarded as perturbations.  For this
purpose, following \cite{Dasgupta:2002hx}, we rescale the fluctuations
and parameters as
\begin{gather}
A   \rightarrow \mu^{-1/2} A   ~,~~~
Y^I \rightarrow \mu^{-1/2} Y^I ~,~~~
C   \rightarrow \mu^{-1/2} C   ~,~~~
\bar{C} \rightarrow \mu^{-1/2} \bar{C} ~,
\notag \\
r_{1,2} \rightarrow \mu r_{1,2} ~,~~~
\epsilon \rightarrow \mu \epsilon ~,~~~ 
t \rightarrow \mu^{-1} t ~.
\label{rescale}
\end{gather}
Under this rescaling, the action $S$ in the background (\ref{b3}) and
(\ref{b6}) becomes
\begin{align}
S =  S_2 + \mu^{-3/2} S_3 + \mu^{-3} S_4 ~,
\label{ssss}
\end{align}
where $S_2$, $S_3$ and $S_4$ do not have $\mu$ dependence.  Now it is
obvious that, in the large $\mu$ limit, $S_3$ and $S_4$ can be treated
as perturbations and the one-loop computation gives the sensible
result.

Based on the structure of (\ref{gb}), we now write the quantum
fluctuations in the $2 \times 2$ block matrix form as follows:
\begin{gather}
A   = \begin{pmatrix}
          0               &   \Phi^0      \\
         \Phi^{0 \dagger} &   0
      \end{pmatrix} ~,~~~
Y^I = \begin{pmatrix}
          0               &   \Phi^I      \\
         \Phi^{I \dagger} &   0
      \end{pmatrix} ~,~~~
\Psi = \begin{pmatrix}
          0             &  \chi \\
         \chi^{\dagger} &  0
       \end{pmatrix} ~,
 \notag \\
C =    \begin{pmatrix}
         0           &  C  \\ 
         C^{\dagger} &  0
       \end{pmatrix} ~,~~~
\bar{C} = \begin{pmatrix}
              0               &  \bar{C}  \\
              \bar{C}^\dagger &  0
           \end{pmatrix} ~.
\label{q-fluct}
\end{gather}
Although we denote the block off-diagonal matrices for the ghosts by
the same symbols with those of the original ghost matrices, there will
be no confusion since $N \times N$ matrices will never appear in what
follows.  The reason why the block-diagonal parts are taken to vanish
is that they do not give any effect on the interaction between fuzzy
spheres but lead to the quantum corrections to each fuzzy spheres.
The issue about the path integral of block-diagonal fluctuations has
been already considered in Refs.~\cite{Sugiyama:2002bw,Shin:2003np},
and it has been shown that each fuzzy sphere does not get quantum
correction at least at one-loop order.

\section{One-loop path integration}
\label{path}

In this section, the path integration of $S_2$ is performed and the
formal expression of the effective action $\Gamma_{\rm eff}$ is
obtained.  The resulting effective action will lead us to have
informations about the interaction between two fuzzy spheres described
by Eqs.~(\ref{b3}) and (\ref{b6}).

The quadratic action $S_2$ may split into three pieces which do not
couple with each other;
\begin{align}
S_2 = \int dt ( L_B + L_G + L_F ) ~,
\label{quad-a}
\end{align}
where $L_G$, $L_G$, and $L_F$ are the Lagrangians for the bosonic,
ghost, and fermionic fluctuations of Eq.~(\ref{q-fluct}) respectively
and their explicit expressions will be presented in due course.  We
note that there is no $\mu$ parameter in $S_2$ due to the rescaling
(\ref{rescale}), and hence the parameter does not appear in the
one-loop effective action.  

Let us begin with defining two useful quantities as
\begin{align}
r \equiv r_2 - r_1 ~,~~~
f(t) = \epsilon^2 + 2 \epsilon r \cos (t/3) ~,
\label{2def}
\end{align}
where $r$ is the mean distance between two fuzzy spheres.  To express
the formal results of path integrations compactly, some propagators
are defined as follows.
\begin{align}
& G_s = \frac{1}{\partial_t^2 + r^2 + \frac{1}{3^2} (j+s)^2}~,~~~
(s=0,~1/2,~1)
 \notag \\
& G_g = \frac{1}{\partial_t^2 + r^2 + \frac{1}{3^2} j(j+1)} ~,
\end{align}
which appear in various places in this section.

In calculating the effective action, the prescription given by Kabat
and Taylor \cite{Kabat:1998im} is usually used.  It is well suited
when the second and higher time derivatives of background can be
ignored.  At the present case, since the background is given by
trigonometric functions, it may not be easy for their prescription to
be adopted directly.  As pointed out in \cite{Shin:2003np}, however, a
simpler formulation is possible when we use the expansion in terms of
the matrix spherical harmonics which has been done
in\cite{Dasgupta:2002hx}.  The basic observation is that the $J^i$
representing the fuzzy sphere satisfies the $SU(2)$ algebra
(\ref{su2}) and the fluctuation matrices in the theory can be regarded
as $SU(2)$ representations.  This leads to the appropriate form of the
Lagrangian which makes the path integration be straightforward.

Let us note that the fluctuation matrices are $N_1 \times N_2$ or $N_2
\times N_1$ ones.  This means that, if we regard an $N_1 \times N_2$
block off-diagonal matrix as an $N_1 N_2$-dimensional reducible
representation of $SU(2)$, it has the decomposition into irreducible
spin $j$ representations with the range $|N_1-N_2|/2 \le j \le
(N_1+N_2)/2-1$, that is, $N_1N_2 =
\bigoplus_{j=|N_1-N_2|/2}^{(N_1+N_2)/2-1} (2j+1)$.  If we denote a
generic fluctuation matrix as $\Phi$, then we have an expansion like
\begin{align}
\Phi = \sum_{j=|N_1-N_2|/2}^{(N_1+N_2)/2-1}
       \sum_{m=-j}^{j}
       \phi_{jm} Y^{N_1 \times N_2}_{jm} ~,
\label{o-exp}
\end{align}
where $Y^{N_1 \times N_2}_{jm}$ is the $N_1 \times N_2$ matrix
spherical harmonics transforming in the irreducible spin $j$
representation and $\phi_{jm}$ is the corresponding spherical mode.
This expansion allows us to write the Lagrangian in terms of the
spherical modes and diagonalize some mass terms which are non-trivial
products of fluctuation matrices.  We will call the resulting
Lagrangian as the {\em diagonalized} Lagrangian.  One should notice
that this does not mean that the diagonalized Lagrangian is the sum of
free Lagrangians of spherical modes.  Since the diagonalization is
matrix diagonalization of mass terms, the Lagrangian may still have
coupling terms between spherical modes even after the diagonalization.
We note that, after the diagonalization, the spherical modes may have
different ranges of $j$ if we keep $-j \le m \le j$.  Since the whole
procedure of diagonalization has been presented in
Refs.~\cite{Dasgupta:2002hx,Shin:2003np} and the diagonalizations in
this paper proceed in exactly the same way, we will not go into any
detail and only present the results.  We also follow the same
notations and conventions of \cite{Shin:2003np} especially for the
spherical modes.

\subsection{Bosonic fluctuation}

Let us first consider the bosonic Lagrangian and evaluate its path
integral.  The Lagrangian is
\begin{align}
L_B = \mathrm{Tr}  
\bigg[ 
& - | \dot{\Phi}^0 |^2 
  + (r^2+f(t)) | \Phi^0 |^2
  + \frac{1}{3^2} \Phi^{0 \dagger} J^i \circ ( J^i \circ \Phi^0 )
 \notag \\
& + | \dot{\Phi}^i |^2 - (r^2+f(t)) | \Phi^i |^2
  - \frac{1}{3^2} | \Phi^i + i \epsilon^{ijk} J^j \circ \Phi^k |^2
  + \frac{1}{3^2} | J^i \circ \Phi^i |^2
 \notag \\
& + | \dot{\Phi}^a |^2 
  - \Big( r^2 + f(t)+\frac{1}{6^2} \Big) | \Phi^a |^2
  - \frac{1}{3^2} \Phi^{a \dagger} J^i \circ ( J^i \circ \Phi^a )
 \notag \\
& + \frac{i}{3} (r+\epsilon) \sin (t/6)
            ( \Phi^{0 \dagger} \Phi^4 - \Phi^{4 \dagger} \Phi^0 )
  - \frac{i}{3} (r-\epsilon) \cos (t/6)
            ( \Phi^{0 \dagger} \Phi^5 - \Phi^{5 \dagger} \Phi^0 ) \,
\bigg] ~.
\label{lb}
\end{align}
Here, adopting the notation of \cite{Dasgupta:2002hx}, we have defined
\begin{equation}
J^i \circ M_{(rs)} \equiv J^i_{(r)} M_{(rs)} - M_{(rs)} J^i_{(s)} ~,
\label{off-com}
\end{equation}
where $M_{(rs)}$ is the $N_r \times N_s$ matrix that is a block at
$r$-th row and $s$-th column in the blocked form of a given matrix
$M$.  In the present case, $r$ and $s$ take values of $1$ and $2$.
For example, if we look at the $2 \times 2$ block matrix form of the
gauge field fluctuation $A$ in Eq.~(\ref{q-fluct}), then
$A_{(ss)}=0$, $A_{(12)}=\Phi^0$, and $A_{(21)}=\Phi^{0
\dagger}$.

The matrix fields, $\Phi^0$, $\Phi^4$, and $\Phi^5$ are coupled with
each other through the fuzzy sphere background.  In order to utilize
the results of our previous work \cite{Shin:2003np} where only the
fuzzy spheres in circular motion have been considered, it is
convenient to define the following matrix variables.
\begin{align}
\Phi^r  \equiv   \cos (t/6) \Phi^4 + \sin (t/6) \Phi^5 ~,~~~
\Phi^\theta 
         \equiv - \sin (t/6) \Phi^4 + \cos (t/6) \Phi^5 ~.
\end{align}
In terms of these fluctuations, the terms in the Lagrangian
(\ref{lb}), which are dependent on $\Phi^4$ and $\Phi^5$, are
rewritten as
\begin{align}
\mathrm{Tr} \bigg[ & | \dot{\Phi}^r |^2 + | \dot{\Phi}^\theta |^2
  - (r^2+f(t)) \left( | \Phi^r |^2 + | \Phi^\theta |^2 \right)
  - \frac{1}{3^2}
    \left(  
           \Phi^{r \dagger} J^i \circ ( J^i \circ \Phi^r )
         + \Phi^{\theta \dagger} J^i \circ ( J^i \circ \Phi^\theta ) 
    \right)
 \notag \\
& + \frac{1}{3} ( \Phi^{r \dagger} \dot{\Phi}^\theta
                   -\Phi^{\theta \dagger} \dot{\Phi}^r )
 - \frac{i}{3} (r - \epsilon \cos (t/3))  ( \Phi^{0 \dagger} \Phi^\theta
                       -\Phi^{\theta \dagger} \Phi^0 ) 
\notag \\
& + \frac{i}{3} \epsilon \sin (t/3) ( \Phi^{0 \dagger} \Phi^r
                       -\Phi^{r \dagger} \Phi^0 ) 
 \bigg] ~.
\end{align}

We observe that, because of the background for the motion in
$x^4$-$x^5$ plane, the transverse $SO(6)$ symmetry is broken down to
$SO(4)$, while the $SO(3)$ symmetry remains intact.  This fact
naturally leads us to break the bosonic Lagrangian (\ref{lb}) into
three parts as follows:
\begin{equation}
L_B = L_{SO(3)} + L_{SO(4)} + L_{\rm rot}~,
\end{equation}
where $L_{SO(3)}$ is the Lagrangian for $\Phi^i$, $L_{SO(4)}$ is for
$\Phi^{a'}$ with $a'=6,7,8,9$, and $L_{\rm rot}$ represents the
rotational part described by $\Phi^r$, $\Phi^\theta$, and the gauge
fluctuation $\Phi^0$.

We first consider $L_{SO(3)}$ and its path integration.  The
diagonalized Lagrangian describes the dynamics of two physical
spherical modes, $\alpha_{jm}$ and $\beta_{jm}$, and one gauge mode,
$\omega_{jm}$.  Its explicit form is
\begin{align}
L_{SO(3)} =
&  \sum^{\frac{1}{2}(N_1+N_2)-2}_{j=\frac{1}{2}|N_1-N_2|-1}
\left[  |\dot{\alpha}_{jm}|^2 
  - \left( r^2 + f(t) + \frac{1}{3^2} ( j+ 1 )^2 \right)
    | \alpha_{jm} |^2 
\right]
  \notag \\
&  + \sum^{\frac{1}{2}(N_1+N_2)}_{j=\frac{1}{2}|N_1-N_2|+1}
\left[  | \dot{\beta}_{jm} |^2
  - \left( r^2 + f(t) + \frac{1}{3^2} j^2 \right) 
    | \beta_{jm} |^2 
\right]
  \notag \\
&  + \sum^{ \frac{1}{2}(N_1+N_2)-1}_{j= \frac{1}{2}|N_1-N_2| } 
\left[
- | \dot{\omega}_{jm} |^2 
+ \left( r^2  + f(t) + \frac{1}{3^2} j (j+1) \right)
   | \omega_{jm} |^2
\right] ~,
\end{align}
where the sum of $m$ over the range $-j \le m \le j$ is implicit. The
path integral of this quadratic Lagrangian is straightforward and
results in
\begin{align}
e^{i \Gamma_{SO(3)}} 
=& e^{i \Gamma_{SO(3)}^{(0)}}
\left(
  \prod^{\frac{1}{2}(N_1+N_2)-2}_{j= \frac{1}{2}|N_1-N_2|-1}
  \left[ \det \left( 1 + G_1 f(t) \right) 
  \right]^{-(2j + 1)}
\right)
\notag \\
\times & 
\left(
  \prod^{\frac{1}{2}(N_1+N_2)}_{j= \frac{1}{2}|N_1-N_2|+1}
  \left[ \det \left( 1 + G_0 f(t) \right) 
  \right]^{-(2j + 1)}
\right)
\notag \\
\times & 
\left(
\prod^{ \frac{1}{2}(N_1+N_2)-1}_{j= \frac{1}{2}|N_1-N_2|} 
\left[
 \det \left( G_g^{-1} + f(t) \right)
\right]^{-(2j + 1)}
\right) ~,
\label{ea-so3}
\end{align} 
where $\Gamma_{SO(3)}^{(0)}$ is the $\epsilon$ independent part from
the path integration of the physical modes $\alpha_{jm}$ and
$\beta_{jm}$, and its explicit form is
\[ e^{i \Gamma_{SO(3)}^{(0)}} = \left(
  \prod^{\frac{1}{2}(N_1+N_2)-2}_{j= \frac{1}{2}|N_1-N_2|-1}
  (\det  G_1^{-1})^{-2j-1} \right) \left(
  \prod^{\frac{1}{2}(N_1+N_2)}_{j= \frac{1}{2}|N_1-N_2|+1}
  (\det G_0^{-1})^{-2j-1} \right)~.
\]  
Since $\Gamma_{SO(3)}^{(0)}$ and other forthcoming
$\epsilon$-independent parts have been already calculated in our
previous work \cite{Shin:2003np}, we do not discuss about them in any
detail but just quote the result that the sum of them vanishes
basically due to supersymmetry, that is,
\begin{align}
\sum_n \Gamma^{(0)}_n = 0 ~,
\end{align}
where $n$ indicates each physical sectors.  The last determinant of
Eq.~(\ref{ea-so3}) coming from the path integral of gauge modes
$\omega_{jm}$ does not split into $\epsilon$ dependent and
independent parts because it is eliminated by ghost contributions
which will be given in the next subsection, that is, Eq.~(\ref{p-og}),
and thus does not contribute to the total effective action
$\Gamma_{\rm eff}$.

Let us turn to the bosonic $SO(4)$ part $L_{SO(4)}$.  The
corresponding spherical modes are $\phi^{a'}_{jm}$ and the
diagonalized Lagrangian is obtained as
\begin{equation}
L_{SO(4)} = \sum^{ \frac{1}{2}(N_1+N_2)-1}_{j= \frac{1}{2}|N_1-N_2|} 
\left[
 | \dot{\phi}^{a'}_{jm} |^2 
- \left( r^2 + f(t) + \frac{1}{3^2} \left( j + \frac{1}{2} \right)^2
  \right) | \phi^{a'}_{jm} |^2
\right] ~,
\end{equation}
where $-j \le m \le j$ and $a'=6,7,8,9$.  Its path integration
leads us to have
\begin{align}
e^{i \Gamma_{SO(4)}} 
= e^{i \Gamma_{SO(4)}^{(0)}}
  \prod^{\frac{1}{2}(N_1+N_2)-1}_{j= \frac{1}{2}|N_1-N_2|}
  \left[ \det \left( 1 + G_{1/2} f(t) \right) 
  \right]^{- 4 (2j + 1)} ~,
\label{ea-so4}
\end{align} 
where $ e^{i \Gamma_{SO(4)}^{(0)}} =
\prod^{\frac{1}{2}(N_1+N_2)-1}_{j= \frac{1}{2}|N_1-N_2|} (\det
G_{1/2}^{-1})^{-2j-1}$.

For the rotational part, the diagonalized Lagrangian describes the
dynamics of three spherical modes, $\phi^0_{jm}$, $\phi^r_{jm}$, and
$\phi^\theta_{jm}$, where the last two are the physical modes and
$\phi^0_{jm}$ is the gauge one.  The Lagrangian for these modes are
\begin{align}
L_{\rm rot} =
 \sum^{ \frac{1}{2}(N_1+N_2)-1}_{j= \frac{1}{2}|N_1-N_2|}  
\Bigg[ 
&- | \dot{\phi}^0_{jm} |^2 
+ \left( r^2  + f(t) + \frac{1}{3^2} j (j+1) \right)
   | \phi^0_{jm} |^2
  \notag \\
& + | \dot{\phi}^r_{jm} |^2 + | \dot{\phi}^\theta_{jm} |^2
- \left( r^2  + f(t) + \frac{1}{3^2} j (j+1) \right)
  \left( | \phi^r_{jm} |^2 + | \phi^\theta_{jm} |^2 \right)
  \notag \\
& + \frac{1}{3} ( \phi^{r*}_{jm} \dot{\phi}^\theta_{jm}
                   -\phi^{\theta *}_{jm} \dot{\phi}^r_{jm} )
  - \frac{i}{3} ( r - \epsilon \cos (t/3) )
       ( \phi^{0*}_{jm} \phi^\theta_{jm}
        -\phi^{\theta *}_{jm} \phi^0_{jm} ) 
 \notag \\
& + \frac{i}{3} \epsilon \sin (t/3) 
       ( \phi^{0*}_{jm} \phi^r_{jm}
        -\phi^{r *}_{jm} \phi^0_{jm} )
\Bigg] ~,
\end{align}
where $-j \le m \le j$.  Although the Lagrangian is quadratic in
fields, the direct path integration is not so easy even at the formal
level because of the explicit time dependent background.  In order to
do the path integration without any ambiguity, we first decouple the
gauge mode from the other modes by performing a field redefinition,
\begin{align*}
\phi^0_{jm} \longrightarrow 
\phi^0_{jm} - \frac{i}{3} G_g \left[ \epsilon \sin (t/3) \phi^r_{jm}
 - (r - \epsilon \cos (t/3) ) \phi^\theta_{jm} \right] ~.
\end{align*}
The modes $\phi^r_{jm}$ and $\phi^\theta_{jm}$ are still coupled with
each other and the terms depending on them contain more complicated
dependence of time dependent background after the above redefinition.
For doing the path integration, we put the terms into the
form of $v^\dagger M v$ where $v$ is a two component column vector
defined as $v= \begin{pmatrix} \phi^r_{jm} \\ \phi^\theta_{jm}
\end{pmatrix}$, and $M$ is the $2 \times 2$ matrix whose components
are determined by the Lagrangian.  Then the path integration of
$\phi^r_{jm}$ and $\phi^\theta_{jm}$ is just that of $v$, which can be
done by applying the following identity to the matrix $M$.
\begin{align}
\begin{pmatrix}
A & B \\ C & D
\end{pmatrix} =
\begin{pmatrix}
A & 0 \\ C & 1
\end{pmatrix}
\begin{pmatrix}
1 & A^{-1} B \\ 0 & D-C A^{-1} B ~.
\end{pmatrix}
\label{miden}
\end{align}
The resulting formal expressions of the path integrations is then
obtained as 
\begin{align}
e^{i \Gamma_{\rm rot}}
=& e^{i \Gamma_{\rm rot}^{(0)}} 
\prod^{ \frac{1}{2}(N_1+N_2)-1}_{j= \frac{1}{2}|N_1-N_2|} 
\Bigg[
\det \left( 1 + G_g f(t) + \frac{\epsilon^2}{3^2} 
       G_g \sin (t/3) \frac{1}{G_g^{-1} + f(t)} \sin (t/3) 
     \right)
\notag \\
\times & \det \left( G_0 G_1 G_g^{-1} K \right)
\det \left( G_g^{-1} + f(t)
      \right)
\Bigg]^{-(2j+1)} ~,
\label{ea-so2}
\end{align}
where $e^{i \Gamma_{\rm rot}^{(0)}} = \prod^{
  \frac{1}{2}(N_1+N_2)-1}_{j= \frac{1}{2}|N_1-N_2|} (\det G_0^{-1}
G_1^{-1})^{-2j-1}$ and we have defined
\begin{align}
K \equiv 
& G_g^{-1} + f(t) 
  + \frac{1}{3^2} (r - \epsilon \cos (t/3) )
   \frac{1}{G_g^{-1} + f(t)} ( r - \epsilon \cos (t/3) )
\notag \\
 + & \frac{1}{3^2} \left( \partial_t 
    - \frac{\epsilon}{3} ( r - \epsilon \cos (t/3) )
\frac{1}{G_g^{-1} + f(t)} \sin (t/3) \right)
\notag \\
 & \times \left( G_g^{-1} + f(t) + \frac{\epsilon^2}{3^2} \sin (t/3)
\frac{1}{G_g^{-1} + f(t)} \sin (t/3) \right)^{-1}
\notag \\
 & \times \left( \partial_t 
    + \frac{\epsilon}{3} \sin (t/3)
\frac{1}{G_g^{-1} + f(t)} ( r - \epsilon \cos (t/3) )  \right) ~.
\end{align}
The last determinant factor in (\ref{ea-so2}) comes from the path
integral of the gauge mode.  We note that $G_0 G_1 G_g^{-1} K = 1 +
\mathcal{O} (\epsilon)$ and thus the second determinant factor in
(\ref{ea-so2}) does not involve non-trivial $\epsilon$ independent
part.

\subsection{Ghost fluctuation}

We turn to the consideration of the path integration for the ghost
part of the action $S_2$ (\ref{quad-a}).  The Lagrangian for the
ghosts is
\begin{align}
L_G =  \mathrm{Tr}
\bigg[ \, 
& \dot{\bar{C}} \dot{C}^{\dagger} 
- (r^2 + f(t)) \bar{C} C^{\dagger}
+ \frac{1}{3^2} (J^i\circ \bar{C})(J^i\circ C^\dagger) \,
\bigg]  
\notag \\
+ \mathrm{Tr} 
\bigg[ \,
& \dot{\bar{C}}^\dagger \dot{C} 
- (r^2 + f(t)) \bar{C}^\dagger C
+ \frac{1}{3^2} (J^i\circ \bar{C}^\dagger)(J^i\circ C) \,
\bigg] \, . 
\end{align}
The spherical modes of $C$ and $\bar{C}$ in the matrix spherical
harmonics expansion are $c_{jm}$ and $\bar{c}_{jm}$ respectively. The
diagonalized Lagrangian is then obtained as
\begin{align}
L_G = \sum^{ \frac{1}{2}(N_1+N_2)-1}_{j= \frac{1}{2}|N_1-N_2|} 
\Bigg[ \,
  \dot{\bar{c}}^*_{jm} \dot{c}^{\dagger}_{jm} 
+ \dot{\bar{c}}^{\dagger *}_{jm} \dot{c}_{jm} 
  - \left( r^2 + f(t)+ \frac{1}{3^2} j(j+1) \right)
  ( \bar{c}^*_{jm} c^{\dagger}_{jm} 
  + \bar{c}^{\dagger *}_{jm} c_{jm} ) \,
\Bigg]\,,
\end{align}
where the sum over $m$ for the range $-j \le m \le j$ is implicit.
The path integral for the above diagonalized Lagrangian is then
immediately evaluated as follows:
\begin{equation}
\prod^{ \frac{1}{2}(N_1+N_2)-1}_{j= \frac{1}{2}|N_1-N_2|} 
\left[
 \det \left( G_g^{-1} + f(t)
      \right)
\right]^{2 (2j + 1)} ~.
\label{p-og}
\end{equation}
As we mentioned earlier, this eliminates the last determinant factors
of (\ref{ea-so3}) and (\ref{ea-so2}) resulting from the path
integrations of the gauge modes.  Thus there is no contribution of the
gauge and ghost modes to the total effective action, as it should be.
Then the physical bosonic effective action $\Gamma_B$ is now given by
\begin{align}
\Gamma_B = \widehat{\Gamma}_{SO(3)} + \Gamma_{SO(4)} 
         + \widehat{\Gamma}_{\rm rot} ~,
\label{physbea}
\end{align}
where $\widehat{\Gamma}_{SO(3)}$ and $\widehat{\Gamma}_{\rm rot}$ are
the same as those of Eqs.~(\ref{ea-so3}) and (\ref{ea-so2}) except for
the absence of the gauge mode contributions.

\subsection{Fermionic fluctuation}

The fermionic Lagrangian is
\begin{align}
L_F =
 2 \mathrm{Tr}
\bigg[ \,
&  i \chi^\dagger \dot{\chi}
 - i \frac{1}{4} \chi^\dagger \gamma^{123} \chi
 + \frac{1}{3} \chi^\dagger \gamma^i J^i \circ \chi
 - r \chi^\dagger
    \left(  \gamma^4 \cos (t/6)
          + \gamma^5 \sin (t/6)
    \right)  \chi 
\notag \\
& - \epsilon \chi^\dagger
     \left(  \gamma^4 \cos (t/6)
           - \gamma^5 \sin (t/6)
     \right)  \chi \,
\bigg] ~.
\label{lf1}
\end{align}
It is convenient to introduce the $SU(2) \times SU(4)$ formulation
since the preserved symmetry in the plane-wave matrix model is
$SO(3)\times SO(6) \sim SU(2)\times SU(4)$ rather than $SO(9)$.  In
this formulation the $SO(9)$ spinor $\chi$ is decomposed as
$\chi_{A\alpha}$ and $\hat{\chi}^{A\beta}$ according to ${\bf 16}
\rightarrow ({\bf 2}, {\bf 4}) + (\bar{\bf 2}, \bar{\bf 4})$, where
$A$ implies a fundamental $SU(4)$ index and $\alpha$ is a fundamental
$SU(2)$ index.  According to this decomposition, we may take the
expression of $\chi$ as
\begin{equation}
\chi = \frac{1}{\sqrt{2}}
\begin{pmatrix}
\chi_{A\alpha} \\ 
\hat{\chi}^A_\alpha
\end{pmatrix} ~,
\label{o-dec}
\end{equation}
where $\hat{\chi}^A_\alpha=\epsilon_{\alpha\beta}
\hat{\chi}^{A\beta}$.  We also rewrite the $SO(9)$ gamma matrices
$\gamma^I$'s in terms of $SU(2)$ and $SU(4)$ ones as follows:
\begin{equation}
\gamma^i = \begin{pmatrix}
               - \sigma^i \times 1 &         0        \\
                         0         & \sigma^i \times 1 
           \end{pmatrix} \,, \quad 
\gamma^a = \begin{pmatrix}
                         0                 & 1\times \rho^a  \\
              1 \times (\rho^a)^{\dagger}  &       0
           \end{pmatrix} \, ,
\end{equation}
where the $\sigma^i$'s are the standard $2 \times 2$ Pauli matrices
and six of $\rho^a$ are taken to form a basis of $4 \times 4$
anti-symmetric matrices. The original $SO(9)$ Clifford algebra
(\ref{so9-cliff}) is satisfied as long as we take normalization so
that the gamma matrices $\rho^a$ with $SU(4)$ indices satisfy the
algebra
\begin{align}
\rho^a \rho^{b \dagger} + \rho^b \rho^{a \dagger} = 2\delta^{ab}\,.
\label{cliff}
\end{align} 

To work in parallel with our previous work \cite{Shin:2003np}, we now
redefine the fermionic field $\hat{\chi}^A_\alpha$ as
\begin{align}
\hat{\chi}^A_\alpha \equiv
  \left[
            (\rho^{4\dagger})^{AB} \cos (t/6)
          + (\rho^{5\dagger})^{AB} \sin (t/6)
  \right]  \tilde{\chi}_{B \alpha} ~,
\end{align}
where we have introduced a new fermionic field $\tilde{\chi}_{A
  \alpha}$ which is in the ${\bf 4}$ of $SU(4)$.  Then, by using the
decomposed form of $\chi$ (\ref{o-dec}) with this redefinition and the
following identities,
\begin{align}
&\left( \rho^4 \cos (t/6) + \rho^5 \sin (t/6) \right)
 \left( \rho^{4\dagger} \cos (t/6) + \rho^{5\dagger} \sin (t/6)
 \right) = 1 ~, 
\notag \\
&\left( \rho^4 \cos (t/6) + \rho^5 \sin (t/6) \right)
 \left( - \rho^{4\dagger} \sin (t/6) + \rho^{5\dagger} \cos (t/6)
      \right) = \rho^4 \rho^{5\dagger} ~,
\end{align}
which are proved via the Clifford algebra (\ref{cliff}), we
may show that the Lagrangian (\ref{lf1}) becomes
\begin{align}
L_F = \mathrm{Tr}
\bigg[
& i \chi^{\dagger A\alpha} \dot{\chi}_{A\alpha} 
  - \frac{1}{4} \chi^{\dagger A\alpha} \chi_{A\alpha}
  - \frac{1}{3} \chi^{\dagger A\alpha}(\sigma^i)_{\alpha}{}^\beta
                J^i\circ \chi_{A\beta} 
\notag \\
& + i\tilde{\chi}^{\dagger A\alpha} \dot{\tilde{\chi}}_{A\alpha} 
  + \frac{1}{4} \tilde{\chi}^{\dagger A\alpha} 
                \tilde{\chi}_{A\alpha}
  + \frac{1}{3} \tilde{\chi}^{\dagger A\alpha}
                (\sigma^i)_{\alpha}{}^\beta
                J^i\circ \tilde{\chi}_{A\beta}
\notag \\
& - \left( r + \epsilon \cos (t/3) \right) 
    \left( \chi^{\dagger A \alpha} \tilde{\chi}_{A \alpha} 
            +\tilde{\chi}^{\dagger A \alpha} \chi_{A \alpha}
      \right)
  + \frac{i}{6} \tilde{\chi}^{\dagger A\alpha}
                ( \rho^4 \rho^{5\dagger} )_A{}^B 
                \tilde{\chi}_{B\alpha}
\notag \\
& -  \epsilon \sin (t/3)
     \left( \chi^{\dagger A \alpha} 
        ( \rho^4 \rho^{5\dagger} )_A{}^B  \tilde{\chi}_{B \alpha} 
        - \tilde{\chi}^{\dagger A \alpha} 
        ( \rho^4 \rho^{5\dagger} )_A{}^B  \chi_{B \alpha}
     \right)
\bigg] \, .
\label{lf3}
\end{align}

For obtaining the diagonalized Lagrangian, we first take the expansion
of $\chi_\alpha$ and $\tilde{\chi}_\alpha$ ($SU(4)$ indices are
suppressed.) in terms of the matrix spherical harmonics like
(\ref{o-exp}) and denote their spherical modes as $(\chi_\alpha)_{jm}$
and $(\tilde{\chi}_\alpha)_{jm}$ respectively.  The diagonalization of
the mass terms for the modes is that for the $SU(2)$ indices and
results in $(\chi_\alpha)_{jm} \rightarrow ( \pi_{jm}, \eta_{jm})$ and
$(\tilde{\chi}_\alpha)_{jm} \rightarrow ( \tilde{\pi}_{jm},
\tilde{\eta}_{jm})$.  The modes $\pi_{jm}$ and $\tilde{\pi}_{jm}$ have
the same $j$-dependent mass of $\frac{1}{3} (j+\frac{3}{4})$ with
$\frac{1}{2}|N_1 - N_2| - \frac{1}{2} \le j \le\frac{1}{2}(N_1 +
N_2)-\frac{3}{2}$.  For the modes $\eta_{jm}$ and $\tilde{\eta}_{jm}$,
their $j$-dependent mass is $\frac{1}{3} (j+\frac{1}{4})$ with
$\frac{1}{2}|N_1 - N_2|+\frac{1}{2} \le j \le\frac{1}{2}(N_1 +
N_2)-\frac{1}{2}$.  All the modes have the same range of $m$ as $-j
\le m \le j$.  Having the spherical modes leading to the
diagonalization for the $SU(2)$ indices, we turn to the consideration
of terms containing $\rho^4 \rho^{5 \dagger}$.  The product $\rho^4
\rho^{5 \dagger}$ measures the $SO(2)$ chirality in the $x^4$-$x^5$
plane where the fuzzy spheres take their motion.  Since $(\rho^4
\rho^{5 \dagger} )^2=-1$, its eigenvalues are $\pm i$.  Each spherical
mode may split into modes having definite $\rho^4 \rho^{5 \dagger}$
eigenvalues as follows:
\begin{align}
&\pi_{jm} = \pi_{+jm} + \pi_{-jm} ~,
&& \eta_{jm} = \eta_{+jm} + \eta_{-jm} ~,
\notag \\
&\tilde{\pi}_{jm}=\tilde{\pi}_{+jm}+\tilde{\pi}_{-jm} ~,
&& \tilde{\eta}_{jm}=\tilde{\eta}_{+jm}+\tilde{\eta}_{-jm} ~,
\label{so2-ch}
\end{align}
where the modes on the right hand sides satisfy
\begin{align}
& \rho^4 \rho^{5 \dagger} \pi_{\pm jm} 
              = \pm i \pi_{\pm jm}~,
& \rho^4 \rho^{5 \dagger} \tilde{\pi}_{\pm jm} 
              = \pm i \tilde{\pi}_{\pm jm}~,
\notag \\
& \rho^4 \rho^{5 \dagger} \eta_{\pm jm} 
              = \pm i \eta_{\pm jm}~,
& \rho^4 \rho^{5 \dagger} \tilde{\eta}_{\pm jm} 
              = \pm i \tilde{\eta}_{\pm jm}~.
\end{align}
After splitting the modes according the $SO(2)$ chirality as in
(\ref{so2-ch}), we now have eight kinds of fermionic spherical modes
in total, in terms of which the fermionic Lagrangian
(\ref{lf3}) is diagonalized.  The resulting Lagrangian is composed of
two independent systems as follows. 
\begin{align}
L_F = L_\pi + L_\eta ~.
\end{align}
The first one $L_\pi$, which we call $\pi$-system, is for $\pi_{\pm
  jm}$ and $\tilde{\pi}_{\pm jm}$, and the other one $L_\eta$, that is
$\eta$-system, for $\eta_{\pm jm}$ and $\tilde{\eta}_{\pm jm}$.  The
Lagrangian for the $\pi$-system is given by
\begin{align}
L_\pi = 
& \sum_{j=\frac{1}{2}|N_1 - N_2| - \frac{1}{2}}^{\frac{1}{2}(N_1
        + N_2)-\frac{3}{2}} 
\Bigg[ \, 
   i \pi_{+jm}^\dagger \dot{\pi}_{+jm} 
  + i \pi_{-jm}^\dagger \dot{\pi}_{-jm}
  - \frac{1}{3} \left(j + \frac{3}{4} \right) 
      (  \pi^\dagger_{+jm} \pi_{+jm} 
       + \pi^\dagger_{-jm} \pi_{-jm} )
\notag \\
&  + i \tilde{\pi}_{+jm}^\dagger \dot{\tilde{\pi}}_{+jm} 
   + \frac{1}{3} \left(j + \frac{1}{4} \right) 
       \tilde{\pi}^\dagger_{+jm} \tilde{\pi}_{+jm}
   + i \tilde{\pi}_{-jm}^\dagger \dot{\tilde{\pi}}_{-jm} 
   + \frac{1}{3} \left(j + \frac{5}{4} \right) 
       \tilde{\pi}^\dagger_{-jm} \tilde{\pi}_{-jm} 
\notag \\
&  -  \left( r + \epsilon e^{i t/3} \right)
      ( \pi^\dagger_{+jm} \tilde{\pi}_{+jm} +
       \tilde{\pi}^\dagger_{-jm} \pi_{-jm})
   -  \left( r + \epsilon e^{-i t/3 } \right)
      (  \pi^\dagger_{-jm} \tilde{\pi}_{-jm} +
        \tilde{\pi}^\dagger_{+jm} \pi_{+jm}) \,
\Bigg] ~,
\label{pi-sys}
\end{align}
and, for the $\eta$-system, we have
\begin{align}
L_\eta = 
& + \sum_{j=\frac{1}{2}|N_1 - N_2| + \frac{1}{2}}^{\frac{1}{2}(N_1
        + N_2)-\frac{1}{2}} 
\Bigg[ \, 
   i \eta_{+jm}^\dagger \dot{\eta}_{+jm} 
  + i \eta_{-jm}^\dagger \dot{\eta}_{-jm}
  - \frac{1}{3} \left(j + \frac{1}{4} \right) 
      (  \eta^\dagger_{+jm} \eta_{+jm} 
       + \eta^\dagger_{-jm} \eta_{-jm} )
\notag \\
&  + i \tilde{\eta}_{+jm}^\dagger \dot{\tilde{\eta}}_{+jm} 
   + \frac{1}{3} \left(j - \frac{1}{4} \right) 
       \tilde{\eta}^\dagger_{+jm} \tilde{\eta}_{+jm}
   + i \tilde{\eta}_{-jm}^\dagger \dot{\tilde{\eta}}_{-jm} 
   +  \frac{1}{3} \left(j + \frac{3}{4} \right) 
       \tilde{\eta}^\dagger_{-jm} \tilde{\eta}_{-jm} 
\notag \\
&  -  \left( r + \epsilon e^{i t/3} \right)
      ( \eta^\dagger_{+jm} \tilde{\eta}_{+jm} 
        + \tilde{\eta}^\dagger_{-jm}  \eta_{-jm})
   -  \left( r + \epsilon e^{-i t/3 } \right)
      (  \eta^\dagger_{-jm} \tilde{\eta}_{-jm} +
        \tilde{\eta}^\dagger_{+jm} \eta_{+jm}) \,
\Bigg] ~,
\label{eta-sys}
\end{align}
where the summation of $m$ over the range $-j \le m \le j$ is
implicit.  

Actually, as one may see from (\ref{pi-sys}) and (\ref{eta-sys}), the
$\pi$-system ($\eta$-system) further splits into two independent
systems as
\begin{align}
L_\pi = L_{\pi_+} + L_{\pi_-} ~,~~~
L_\eta = L_{\eta_+} + L_{\eta_-} ~,
\end{align}
where $L_{\pi_\pm}$ ($L_{\eta_\pm}$) describes the dynamics of
$\pi_{\pm jm}$ and $\tilde{\pi}_{\pm jm}$ ($\eta_{\pm jm}$ and
$\tilde{\eta}_{\pm jm}$).  Thus we have four independent systems in
the fermionic sector and are ready to do path integration.  By the
way, we note that all the four systems have exactly the same
structure.  This leads us to consider a prototypical system and its
path integration instead of treating the whole system at once.  Let us
call such a prototypical system the $\psi$-system. Then its Lagrangian
may be given by
\begin{align}
L_\psi =  \sum_j \Big[
& i \psi^\dagger_{jm} \dot{\psi}_{jm} + m_j \psi_{jm}^\dagger \psi_{jm}
  + i \tilde{\psi}_{jm}^\dagger \dot{\tilde{\psi}}_{jm} 
  + \tilde{m}_j \tilde{\psi}_{jm}^\dagger \tilde{\psi}_{jm}
\notag \\
& - (r + \epsilon e^{i t/3} ) \psi_{jm}^\dagger \tilde{\psi}_{jm}
  - (r + \epsilon e^{- i t/3} ) \tilde{\psi}_{jm}^\dagger \psi_{jm}
\Big] ~,
\end{align}
which has obviously the same structure with that of $L_{\pi_\pm}$ and
$L_{\eta_\pm}$.  The range of $j$ is specified according to which
system among $L_{\pi_\pm}$ and $L_{\eta_\pm}$ we relate to the
$\psi$-system and the sum of $m$ over the range $-j \le m \le j$ is
implicit.  The fermions $\psi_{jm}$ and $\tilde{\psi}_{jm}$ should be
two component fermions since the spherical modes in (\ref{pi-sys}) and
(\ref{eta-sys}) have two components (Although the spherical modes have
four components since each of them has a fundamental $SU(4)$ index,
the $SO(2)$ chirality projection reduces the number of independent
components from four to two.).  The path integration can be done by
applying the identity (\ref{miden}) and the formal expression of the
effective action for the $\psi$-system is then obtained as
\begin{align}
e^{i \Gamma_\psi} = e^{i \Gamma_\psi^{(0)}} \prod_j \left[ \det \left( 1 
 - \frac{1}{( i \partial_t + m_j )( i \partial_t + \tilde{m}_j )-r^2} E
\right) \right]^{2(2j+1)} ~,
\label{eff-proto}
\end{align}
where $e^{i \Gamma_\psi^{(0)}} = \prod_j \left[ \det (( i \partial_t +
  m_j )( i \partial_t + \tilde{m}_j )-r^2 ) \right]^{2 (2j+1)}$ and
$E$ is defined as
\begin{align}
E \equiv \epsilon r e^{i t/3} 
+ \epsilon r ( i \partial_t + m_j ) e^{-i t/3} 
  \frac{1}{ i \partial_t + m_j} 
+ \epsilon^2 ( i \partial_t + m_j ) e^{-i t/3} 
  \frac{1}{ i \partial_t + m_j} e^{i t/3} ~.
\end{align}

If we now denote the effective actions obtained after the path
integrations for the systems $L_{\pi_\pm}$ and $L_{\eta_\pm}$ as
$\Gamma_{\pi_\pm}$ and $\Gamma_{\eta_\pm}$ respectively, their formal
expressions are obtained from (\ref{eff-proto}) by substituting the
data of each system for those of the $\psi$-system.  The detailed
correspondence between each system and the $\psi$-system is listed in
Table \ref{t1}.  Having the effective action for each system, we get
the full fermionic effective action $\Gamma_F$ as follows.
\begin{align}
\Gamma_F = \Gamma_\pi + \Gamma_\eta ~,
\label{physfea}
\end{align}
where $\Gamma_\pi = \Gamma_{\pi_+} + \Gamma_{\pi_-}$ and $\Gamma_\eta
= \Gamma_{\eta_+} + \Gamma_{\eta_-}$.

\begin{table}
\begin{center}
\setlength{\extrarowheight}{2mm}
\begin{tabular}{ccccc}
\hline
$L_\psi$ & $\psi_{jm}$ & $\tilde{\psi}_{jm}$ 
                    & $m_j$ & $\tilde{m}_j$ \\[2mm]
\hline
$L_{\pi_+}$ & $\pi_{+jm}$  & $\tilde{\pi}_{+jm}$ & 
    $-\frac{1}{3}(j+\frac{3}{4})$ & $\frac{1}{3}(j+\frac{1}{4})$
        \\[2mm]
$L_{\pi_-}$ & $\tilde{\pi}_{-jm}$  & $\pi_{-jm}$ & 
    $\frac{1}{3}(j+\frac{5}{4})$ & $-\frac{1}{3}(j+\frac{3}{4})$
        \\[2mm]
$L_{\eta_+}$ & $\eta_{+jm}$  & $\tilde{\eta}_{+jm}$ & 
    $-\frac{1}{3}(j+\frac{1}{4})$ & $\frac{1}{3}(j-\frac{1}{4})$
        \\[2mm]
$L_{\eta_-}$ & $\tilde{\eta}_{-jm}$  & $\eta_{-jm}$ & 
    $\frac{1}{3}(j+\frac{3}{4})$ & $-\frac{1}{3}(j+\frac{1}{4})$
        \\[2mm]
\hline
\end{tabular}
\end{center}
\caption{Correspondence of each fermion system with the 
prototypical $\psi$-system}
\label{t1}
\end{table}

\section{Effective action}
\label{eff-int}

We have obtained the formal expressions of the effective actions for
various sectors.  In this section, through the explicit calculation,
we give the one-loop effective action of the plane-wave matrix model
in the fuzzy sphere background described by (\ref{b3}) and (\ref{b6}).

The determinant factors appearing in the formal expressions of the
effective actions in the last section contain time dependent
backgrounds.  This fact makes the exact evaluation of the one-loop
effective action too difficult.  In this case, it is necessary to find
a certain perturbation parameter and evaluate each determinant factor
by expanding it in terms of the parameter.  As for our problem, one
may notice that $\epsilon$ can be taken to be such a perturbative
expansion parameter.  That is to say, we get, as an example, an
expansion like
\begin{align}
\det (1+\epsilon g) = 
   \exp \left( \epsilon {\rm tr} g 
  - \frac{1}{2} \epsilon^2 {\rm tr} g^2  + \dots \right) ~,
\end{align}
where ${\rm tr}$ is the functional trace, $g$ is an operator and an
identity $\det M = \exp ({\rm tr} \ln M)$ has been used.  The
perturbative expansion of determinant factors in terms of $\epsilon$
leads to the same type of expansion of the one-loop effective action
as
\begin{align}
\Gamma_{\rm eff} 
= \Gamma_{\rm eff}^{(0)} + \epsilon^2 \Gamma_{\rm eff}^{(2)}
          + \epsilon^4 \Gamma_{\rm eff}^{(4)} 
          + \mathcal{O} ( \epsilon^6 ) ~.
\label{gexpand}
\end{align}
The reason why we do not have the terms with odd powers of $\epsilon$
will be explained a little bit later.  The coefficient
$\Gamma^{(n)}_{\rm eff}$ at the $\epsilon^n$-order is a function of
$r$, $N_1$, and $N_2$, and is given by
\begin{align}
\Gamma_{\rm eff}^{(n)} =
\Gamma_B^{(n)} + \Gamma_F^{(n)} ~,
\end{align}
where $\Gamma_B^{(n)}$ and $\Gamma_F^{(n)}$ are the expansion
coefficients at the $\epsilon^n$-order in the $\epsilon$ expansions of
the bosonic and fermionic effective actions, (\ref{physbea}) and
(\ref{physfea}), respectively.

The expansion parameter $\epsilon$ is basically related to the change
of radial distance between two fuzzy spheres in the evolution of time.
In this sense, $\epsilon$ is very similar to the relative velocity $v$
between two gravitons in the flat matrix model calculation of two
graviton scattering amplitude \cite{Banks:1997vh}.  As for the flat
matrix model, the nontrivial interaction begins at the $v^4$ order.
This makes us expect that non-vanishing contributions to
$\Gamma_{\rm eff}$ start also at the $\epsilon^4$ order.  As we will
see, it is indeed so.  Thus what we are interested in is the
evaluations of $\Gamma^{(n)}_{\rm eff}$ up to $n=4$.

The structure of $\Gamma^{(n)}_{\rm eff}$ is simply a sum of
functional traces of several propagators and trigonometric functions.
Each functional trace can be evaluated in the momentum space
representation by noting that
\begin{align}
\langle \omega | \cos ( a t ) | \omega' \rangle
 =& \frac{1}{2} [ \delta (\omega-\omega' +a)
                + \delta (\omega-\omega' -a) ] ~,
\notag \\
\langle \omega | \sin ( a t ) | \omega' \rangle
 =& \frac{1}{2i} [ \delta (\omega-\omega' +a)
                - \delta (\omega-\omega' -a) ] ~,
\label{tft}
\end{align}
where the symbol $\omega$ is introduced as the conjugate momentum of
time $t$ and $a=1/3$ in our case.  We note that each trigonometric
function is associated with an $\epsilon$ parameter, as one may see
from the formal expressions of path integrations, (\ref{ea-so3}),
(\ref{ea-so4}), (\ref{ea-so2}), and (\ref{eff-proto}).  Due to this
fact, in the $\epsilon$ expansion of the effective action, the
functional traces contain even (odd) number of trigonometric functions
at the $\epsilon^n$ order with even (odd) $n$.  For odd $n$, according
to Eq.~(\ref{tft}), there are odd number of $\delta$ functions in the
momentum space calculation of each functional trace.  A bit of
manipulation shows that we always end up with vanishing result because
it involves single $\delta$ function whose argument is non-zero
number.  Thus all the functional traces appearing at the $\epsilon^n$
order with odd $n$ vanish and we get $\Gamma^{(n)}_{\rm eff} = 0$ for
odd $n$.  This explains the reason why there are no terms with odd
powers of $\epsilon$.  For even $n$, on the other hand, the
calculation of functional traces lead to results involving $\delta (0)$
which transforms into $\int dt$ in the position space representation.
Thus we may have non-vanishing contributions to the effective action
for even $n$.  

As an additional remark for the contributions at the even powers of
$\epsilon$, we note that they do not depend on the sign of the average
distance $r$ between two fuzzy spheres.  From the definition of $r$,
(\ref{2def}), one may argue that the effective action get the sign
change when $r_1 > r_2$.  However, by looking at the formal
expressions (\ref{ea-so3}), (\ref{ea-so4}), (\ref{ea-so2}), and
(\ref{eff-proto}), it is easy to see that changing the sign of $r$ is
completely equivalent to changing the sign of $\epsilon$.  This means
that the results obtained at the even powers of $\epsilon$ is
independent from the sign of $r$.  Therefore, from now on, $r$ may be
understood as $|r|$.

In evaluating $\Gamma^{(n)}_{\rm eff}$ up to $n=4$, we should compute
a large number of functional traces; it is over 1400.  However, there
is no need to compute all of them, since we focus on the long distance
dynamics of fuzzy spheres.  The result of each functional trace is
basically a complicated combination of the form such as $1/(r^2+j^2)$.
Since the maximum value of $j$ is roughly $N$, the long distance
expansion is valid when $r \gg N$.  In this paper, we consider the
expansion up to the $1/r^9$ order.  For a given functional trace, the
leading power of $r$ is evaluated by simple power counting without
explicit computation of the trace.  This helps us to reduce the number
of functional traces.  Up to the order of interest, the reduced number
is about 900.  However, it is still a large number and thus it is not
practically easy to do the computation by hand.  In this situation, it
is natural to adopt the computer algebra system such as the
Mathematica \cite{wolfram}.  Actually we have performed all the
calculations by computer.  Since however the results at the
intermediate stage require huge amount of space for their
presentation, we will just give final results in what follows.

The $n=0$ case corresponds to the fuzzy spheres in circular motion and
has been already considered in our previous work \cite{Shin:2003np},
where we have shown that
\begin{align}
\Gamma^{(0)}_{\rm eff} = 0 ~,
\label{ge0eff}
\end{align}
due to a non-trivial cancellation between bosonic and fermionic
contributions.  We note that this result is valid for all values of
$r$ and is thus exact at the one-loop level.  Single fuzzy sphere in
circular motion is known to be a BPS object.  The vanishing effective
action for the fuzzy spheres in circular motion means that the
configurations composed of two or more fuzzy spheres in circular
motion are also supersymmetric and BPS objects.

For $n=2$, as one sees in Eq.~(\ref{ge2}), there is again non-trivial
cancellation $ \Gamma^{(2)}_B = -\Gamma^{(2)}_F$, which leads to
\begin{align}
\Gamma^{(2)}_{\rm eff} = 0 ~.
\label{ge2eff}
\end{align}
At first glance, this seems to be an approximate result since the
cancellation has been seen up to the $1/r^9$ order.  However, the
numerical analysis of $\Gamma^{(2)}_{\rm eff}$ before taking the long
distance limit shows that (\ref{ge2eff}) is true with a great
accuracy.  Thus we may argue that the cancellation is exact at least
at the one-loop level.

The non-vanishing contributions start to appear at the quartic order in
$\epsilon$.  The bosonic as well as the fermionic parts, (\ref{ge4b})
and (\ref{ge4f}), have the $1/r^3$ and $1/r^5$ type interactions at
the leading and the next-to-leading orders.  However, there are
cancellations for these interactions and hence the $1/r^7$ becomes the
actual leading interaction in $\Gamma^{(4)}_{\rm eff}$, which is
\begin{align}
\Gamma_{\rm eff}^{(4)} = \int dt \left[
 \frac{35}{2^7 \cdot 3} \frac{N_1 N_2}{r^7}
-\frac{385}{2^{11} \cdot 3^3} 
  \big[ 2 (N_1^2 + N_2^2) -1 \big] \frac{N_1 N_2}{r^9} \right]
+  \mathcal{O} ( r^{-11} ) ~.
\label{ge4eff}
\end{align}

The total one-loop effective action (\ref{gexpand}) up to the
$\epsilon^4$-order is then obtained as
\begin{align}
\Gamma_{\rm eff} 
= \epsilon^4 \Gamma_{\rm eff}^{(4)} + \mathcal{O} ( \epsilon^6 ) ~,
\end{align}
where $\Gamma_{\rm eff}^{(4)}$ is that of Eq.~(\ref{ge4eff}).  The
fact that the leading order interaction in the long distance limit is
given by the form of $1/r^7$ clearly shows that the two fuzzy spheres
interpreted as giant gravitons really behave as gravitons.  Beyond the
leading order interaction showing the attraction between fuzzy
spheres, there is also a sub-leading order interaction.  Interestingly,
it has the sign that is opposite from that of the leading one.  This
indicates an interesting possibility about the bound state of fuzzy
spheres.  If we simply take $N_1=N_2=N$, then, from
Eq.~(\ref{ge4eff}), we may estimate roughly the size $r_B$ of the
bound state as $r_B \sim 0.6 N$.  Unfortunately, however, it is
difficult to trust this result, since the long distance limit is valid
only when $r \gg N$.  In order to answer the question about the bound
state, it is necessary to get the full expression at least at the
$\epsilon^4$ order.  At present, the bound state problem is open.

\section*{Acknowledgments}
The work of H.S. was supported by Korea Science and Engineering
Foundation (KOSEF).  The work of K.~Y.\ is supported in part by JSPS
Research Fellowships for Young Scientists.

\appendix
\section{Effective actions from various sectors}
\label{appen}

We present the resulting formula for the effective actions obtained
from the various sectors.  As mentioned in a previous section, each
effective action is expanded in even powers of the parameter
$\epsilon$ as follows.
\begin{align}
\Gamma = \Gamma^{(0)} + \epsilon^2 \Gamma^{(2)}
          + \epsilon^4 \Gamma^{(4)} 
          + \mathcal{O} ( \epsilon^6 ) ~,
\end{align}
where $\Gamma^{(n)}$ is the effective action at the $\epsilon^n$
order.  The results are given up to quartic order in $\epsilon$, and,
at each order of $\epsilon$, we consider the expansion up to the
$r^{-9}$ order in the long distance (large $r$) limit.

Since the effective action at the $\epsilon^0$-order has been already
given in \cite{Shin:2003np}, we turn to the $\epsilon^2$-order.  At
this order, the effective actions obtained from the bosonic and
fermionic sectors are equal up to overall sign;
\begin{align}
\Gamma_B^{(2)}  = - \Gamma_F^{(2)} = 
&\int dt \bigg[
  -2 \frac{N_1 N_2}{r}
  -\frac{1}{2^3 \cdot 3^2} 
   \big[ 2 (N_1^2 +N_2^2) + 9 \big] \frac{N_1 N_2}{r^3}
\notag \\
& + \frac{1}{2^7 \cdot 3^4} \big[ 18 (N_1^4+N_2^4) 
    + 110 (N_1^2+N_2^2) + 60 N_1^2 N_2^2 - 7 \big]
     \frac{N_1 N_2}{r^5}
\notag \\
& - \frac{1}{2^{10} \cdot 3^7} \big[ 150 (N_1^6+N_2^6) 
     + 1365 (N_1^4+N_2^4) - 539 (N_1^2+N_2^2)
\notag \\
&    + 1050 N_1^2 N_2^2 (N_1^2 + N_2^2) + 4550 N_1^2 N_2^2
     - 505 \big]  \frac{N_1 N_2}{r^7}
\notag \\
& + \frac{1}{2^{15} \cdot 3^8} \big[ 490 (N_1^8+N_2^8) 
     + 6300 (N_1^6+N_2^6) - 5922 (N_1^4 + N_2^4)
\notag \\ 
&    - 8292 (N_1^2+N_2^2) + 5880 N_1^2 N_2^2 (N_1^4  + N_2^4) 
      + 44100 N_1^2 N_2^2 (N_1^2+N_2^2)
\notag \\
&   + 12348 N_1^4 N_2^4 -19740 N_1^2 N_2^2+8901 \big]  
   \frac{N_1 N_2}{r^9} \bigg] + \mathcal{O} ( r^{-11} ) ~.
\label{ge2}
\end{align}
Thus there is no contribution to the total effective action.

At the $\epsilon^4$-order, we get non-trivial results.  To show the
non-triviality in more detail, we split the effective actions from the
bosonic and fermionic sectors as follows.
\begin{align}
\Gamma_B^{(4)} = \widehat{\Gamma}_{SO(3)}^{(4)} + \Gamma_{SO(4)}^{(4)} 
                + \widehat{\Gamma}_{\rm rot}^{(4)} ~,~~~
\Gamma_F^{(4)} = \Gamma_\pi^{(4)} + \Gamma_\eta^{(4)} ~.
\end{align}
We would like to note that, as mentioned in (\ref{physbea}),
$\widehat{\Gamma}_{SO(3)}^{(4)}$ and $\widehat{\Gamma}_{\rm
  rot}^{(4)}$ contain the contributions only from the physical modes.

As for the bosonic sector, the three sub-sectors contribute as
\begin{align}
\widehat{\Gamma}_{SO(3)}^{(4)}=
&\int dt \bigg[ 
  -\frac{1}{32} \frac{N_1 N_2}{r^3}
  -  \frac{1}{2^8 \cdot 3} \big[ 3 (N_1^2 +  N_2^2) - 8 \big] 
  \frac{N_1 N_2}{r^5}
\notag \\
& + \frac{1}{2^{12} \cdot 3^4} 
 \big[ 225 (N_1^4 + N_2^4) 
  + 1000 (N_1^2 + N_2^2)
  + 750 N_1^2 N_2^2 - 4464 \big] \frac{N_1 N_2}{r^7}
\notag \\
& -  \frac{1}{2^{15} \cdot 3^7}
 \big[ 3675 (N_1^6+N_2^6) + 47040 (N_1^4+N_2^4)- 290864(N_1^2+N_2^2) 
\notag \\
& \hspace{18mm} 
 +25725 N_1^2 N_2^2 (N_1^2+N_2^2) + 156800 N_1^2 N_2^2 + 302432 \big]
  \frac{N_1 N_2}{r^9} \bigg]
\notag \\
& + \mathcal{O} ( r^{-11} ) ~,
\end{align}
\begin{align}
\Gamma_{SO(4)}^{(4)}= 
&\int dt \bigg[
  -\frac{1}{16} \frac{N_1 N_2}{r^3}
  -  \frac{1}{2^7 \cdot 3} \big[ 3 (N_1^2 +  N_2^2) - 23 \big] 
  \frac{N_1 N_2}{r^5}
\notag \\
& + \frac{1}{2^{11} \cdot 3^4} 
 \big[ 225 (N_1^4 + N_2^4) 
  -2150 (N_1^2 + N_2^2)
  + 750 N_1^2 N_2^2 + 3661 \big] \frac{N_1 N_2}{r^7}
\notag \\
& -  \frac{1}{2^{14} \cdot 3^7}
 \big[ 3675 (N_1^6+N_2^6) -52185 (N_1^4+N_2^4)+ 208201(N_1^2+N_2^2) 
\notag \\
& \hspace{18mm} 
 +25725 N_1^2 N_2^2 (N_1^2+N_2^2) - 173950 N_1^2 N_2^2 - 206923 \big]
  \frac{N_1 N_2}{r^9} \bigg]
\notag \\
& + \mathcal{O} ( r^{-11} ) ~,
\end{align}
\begin{align}
\widehat{\Gamma}_{\rm rot}^{(4)}=
& \int dt \bigg[  
  -\frac{1}{32} \frac{N_1 N_2}{r^3}
  -  \frac{1}{2^8 \cdot 3} \big[ 3 (N_1^2 +  N_2^2) + 220 \big] 
  \frac{N_1 N_2}{r^5}
\notag \\
& + \frac{1}{2^{12} \cdot 3^4} 
 \big[ 225 (N_1^4 + N_2^4) 
  + 16000 (N_1^2 + N_2^2)
  + 750 N_1^2 N_2^2 + 21296 \big] \frac{N_1 N_2}{r^7}
\notag \\
& -  \frac{1}{2^{15} \cdot 3^7}
 \big[ 3675 (N_1^6+N_2^6) + 320460 (N_1^4+N_2^4)+586096(N_1^2+N_2^2) 
\notag \\
& \hspace{18mm} 
 +25725 N_1^2 N_2^2 (N_1^2+N_2^2) +1068200 N_1^2 N_2^2 -372928 \big]
  \frac{N_1 N_2}{r^9} \bigg]
\notag \\
& + \mathcal{O} ( r^{-11} ) ~.
\end{align}
These three contributions compose the bosonic effective action, which
is
\begin{align}
\Gamma_B^{(4)} = 
& \int dt \bigg[ 
-\frac{1}{8} \frac{N_1 N_2}{r^3} 
-  \frac{1}{2^7 \cdot 3} \big[ 6 (N_1^2 +  N_2^2) + 83 \big] 
  \frac{N_1 N_2}{r^5}
\notag \\
& + \frac{1}{2^{11} \cdot 3^4} 
 \big[ 450 (N_1^4 + N_2^4) 
  + 6350 (N_1^2 + N_2^2)
  + 1500 N_1^2 N_2^2 +12077 \big] \frac{N_1 N_2}{r^7}
\notag \\
& -  \frac{1}{2^{14} \cdot 3^7}
 \big[ 7350 (N_1^6+N_2^6) +131565 (N_1^4+N_2^4)+355817(N_1^2+N_2^2) 
\notag \\
& \hspace{18mm} 
 +51450 N_1^2 N_2^2 (N_1^2+N_2^2) +438550 N_1^2 N_2^2 -242171 \big]
  \frac{N_1 N_2}{r^9} \bigg]
\notag \\
& + \mathcal{O} ( r^{-11} ) ~.
\label{ge4b}
\end{align}

The fermionic sector has two sub-sectors, which give the following
contributions;
\begin{align}
\Gamma_\pi^{(4)} = 
& \int dt \bigg[ 
 \frac{1}{16}(N_1 - 1) \frac{N_2}{r^3} 
+ \frac{1}{2^8 \cdot 3} \big[ 6 N_1^3 + N_1 ( 6 N_2^2 + 83 ) 
   - 12 N_1^2 - 4 N_2^2 -79 \big] \frac{N_2}{r^5}
\notag \\
& -\frac{1}{2^{12} \cdot 3^5}
  \big[ 1350 N_1^5 + 150 N_1^3 ( 30 N_2^2 + 127)
    + 3 N_1 ( 450 N_2^4 + 6350 N_2^2 - 3043 )
\notag \\
&   - 4050 N_1^4 - 150 N_1^2 (54 N_2^2 + 215) - 810 N_2^4
    - 10750 N_2^2 + 19789 \big] \frac{N_2}{r^7}
\notag \\
& +\frac{1}{2^{15} \cdot 3^7} 
  \big[ 7350 N_1^7 + 735 N_1^5 ( 70 N_2^2 + 179 )
   + 7 N_1^3 ( 7350 N_2^4 + 62650 N_2^2 - 20449)
\notag \\
&  + N_1 ( 7350 N_2^6 + 131565 N_2^4 - 143143 N_2^2 + 7309 )
   - 29400 N_1^6 
\notag \\
& - 735 N_1^4 ( 200 N_2^2 + 407) 
  - 21 N_1^2 ( 4200 N_2^4 + 28490 N_2^2 - 33431 )
\notag \\
& - 4200 N_2^6 - 59829 N_2^4 + 234017 N_2^2 - 250307 
  \big] \frac{N_2}{r^9} \bigg]
  + \mathcal{O} ( r^{-11} ) ~,
\end{align}
\begin{align}
\Gamma_\eta^{(4)} = 
& \int dt \bigg[
\frac{1}{16}(N_1 + 1) \frac{N_2}{r^3} 
+ \frac{1}{2^8 \cdot 3} \big[ 6 N_1^3 + N_1 ( 6 N_2^2 + 83 ) 
   + 12 N_1^2 + 4 N_2^2 + 79 \big] \frac{N_2}{r^5}
\notag \\
& -\frac{1}{2^{12} \cdot 3^5}
  \big[ 1350 N_1^5 + 150 N_1^3 ( 30 N_2^2 + 127)
    + 3 N_1 ( 450 N_2^4 + 6350 N_2^2 - 3043 )
\notag \\
&   + 4050 N_1^4 + 150 N_1^2 (54 N_2^2 + 215) + 810 N_2^4
    + 10750 N_2^2 - 19789 \big] \frac{N_2}{r^7}
\notag \\
& +\frac{1}{2^{15} \cdot 3^7} 
  \big[ 7350 N_1^7 + 735 N_1^5 ( 70 N_2^2 + 179 )
   + 7 N_1^3 ( 7350 N_2^4 + 62650 N_2^2 - 20449)
\notag \\
&  + N_1 ( 7350 N_2^6 + 131565 N_2^4 - 143143 N_2^2 + 7309 )
   + 29400 N_1^6 
\notag \\
& + 735 N_1^4 ( 200 N_2^2 + 407) 
  + 21 N_1^2 ( 4200 N_2^4 + 28490 N_2^2 - 33431 )
\notag \\
& + 4200 N_2^6 + 59829 N_2^4 - 234017 N_2^2 + 250307 
  \big] \frac{N_2}{r^9} \bigg]
  + \mathcal{O} ( r^{-11} ) ~.
\end{align}
The resulting fermionic effective action is then given by
\begin{align}
\Gamma_F^{(4)} = 
& \int dt \bigg[
 \frac{1}{8} \frac{N_1 N_2}{r^3} 
+  \frac{1}{2^7 \cdot 3} \big[ 6 (N_1^2 +  N_2^2) + 83 \big] 
  \frac{N_1 N_2}{r^5}
\notag \\
& - \frac{1}{2^{11} \cdot 3^4} 
 \big[ 450 (N_1^4 + N_2^4) 
  + 6350 (N_1^2 + N_2^2)
  + 1500 N_1^2 N_2^2 -3043 \big] \frac{N_1 N_2}{r^7}
\notag \\
& +  \frac{1}{2^{14} \cdot 3^7}
 \big[ 7350 (N_1^6+N_2^6) +131565 (N_1^4+N_2^4)-143143 (N_1^2+N_2^2) 
\notag \\
& \hspace{18mm} 
 +51450 N_1^2 N_2^2 (N_1^2+N_2^2) +438550 N_1^2 N_2^2 +7309 \big]
  \frac{N_1 N_2}{r^9} \bigg]
\notag \\
& + \mathcal{O} ( r^{-11} ) ~.
\label{ge4f}
\end{align}

\end{document}